\documentclass[12pt,preprint]{aastex}

\usepackage{epsfig}
\usepackage{amsmath}
\usepackage{amssymb}
\usepackage{amsbsy}
\usepackage{placeins}
\usepackage{color}
\usepackage{lscape}

\newcommand{\D}{\mbox{d}}

\newcommand{\iles}{{\sc iles}}

\newcommand{\Ka}{\mathrm{Ka}}
\newcommand{\Da}{\mathrm{Da}}

\usepackage{ulem}
\definecolor{green2}{rgb}{0,0.9,0}

\newcommand{\gtaprx}{\lower .1ex\hbox{\rlap{\raise .6ex\hbox{\hskip .3ex
        {\ifmmode{\scriptscriptstyle >}\else
                {$\scriptscriptstyle >$}\fi}}}
        \kern -.4ex{\ifmmode{\scriptscriptstyle \sim}\else
                {$\scriptscriptstyle\sim$}\fi}}}
\newcommand{\ltaprx}{\lower .1ex\hbox{\rlap{\raise .6ex\hbox{\hskip .3ex
        {\ifmmode{\scriptscriptstyle <}\else
                {$\scriptscriptstyle <$}\fi}}}
        \kern -.4ex{\ifmmode{\scriptscriptstyle \sim}\else
                {$\scriptscriptstyle\sim$}\fi}}}

\usepackage{epsf,color,dcolumn}

\newcolumntype{d}{D{.}{.}{-1}}

\epsfverbosetrue

\begin{document}

\title{Turbulent Oxygen Flames in Type Ia Supernovae}

\author{A.~J.~Aspden\altaffilmark{1}, J.~B.~Bell\altaffilmark{1}, and
  S.~E.~Woosley\altaffilmark{2}}

\altaffiltext{1}{Lawrence Berkeley National Laboratory, 1 Cyclotron
Road, MS 50A-1148, Berkeley, CA 94720}
\altaffiltext{2}{Department of Astronomy and Astrophysics, University
of California at Santa Cruz, Santa Cruz, CA 95064}

\begin{abstract}
In previous studies, we examined turbulence-flame interactions in
carbon-burning thermonuclear flames in Type Ia
supernovae.  In this study, we consider turbulence-flame interactions
in the trailing oxygen flames.  The two aims of the paper are to
examine the response of the inductive oxygen flame to intense levels
of turbulence, and to explore the possibility of transition to 
detonation in the oxygen flame.
Scaling arguments analogous to the carbon flames are presented and
then compared against three-dimensional simulations for
a range of Damk\"ohler numbers ($\Da_{16}$) at a fixed Karlovitz
number.  The simulations suggest that turbulence does not significantly
affect the oxygen flame when $\Da_{16}<1$, and the flame burns inductively
some distance behind the carbon flame.  However, for $\Da_{16}>1$, 
turbulence enhances heat transfer and drives the propagation of a flame
that is {\em narrower} than the corresponding inductive flame would be.
Furthermore, burning under these conditions appears to occur as part of a 
combined carbon-oxygen turbulent flame with complex compound structure.
The simulations do not appear to support the possibility of a transition
to detonation in the oxygen flame, but do not preclude it either.
\end{abstract}

\keywords{supernovae: general --- white dwarfs --- hydrodynamics ---
          nuclear reactions, nucleosynthesis, abundances --- conduction ---
          methods: numerical --- turbulence --- distributed flames}

\section{INTRODUCTION}
\label{sec:intro}

A major uncertainty in the modeling of Type Ia supernovae
(SN Ia) is the physical process whereby a subsonic deflagration
transitions to a detonation. Such a transition seems to be
required by the observations \citep{Kozma05,Hoflich95,Mazzali07,Kasen09}, at
least within the context of the popular ``single degenerate'' Chandrasekhar-mass
model. Previous papers \citep[e.g.][]{Khokhlov1997,niemeyerwoosley1997,Woosley09,Aspden10} have
focused on the possibility of a transition to detonation in
carbon-rich material as the deflagration enters the ``distributed
burning regime'' where burning becomes slow enough for turbulence to
disrupt the flame, mixing hot ash and cold fuel. Other papers
\citep[e.g.][]{Plewa04} have explored the possibility that detonation
may be mechanically induced by the collision of burning waves near
the surface of the white dwarf. The present situation is
inconclusive.  The collisions may not be strong enough to robustly
cause a detonation \citep{Roepke07} and the amount of turbulence
required for a spontaneous detonation in the distributed regime is
quite large \citep{Woosley09}.

Here we consider a third possibility - that the necessary carbon
detonation actually begins as an {\sl oxygen} detonation in a hybrid
flame. This possibility has been recently considered \citep{Woosley10} in
a one-dimensional study. Carbon burning produces oxygen-rich ash
that still contains a large potential reservoir of nuclear energy. The
oxygen ash is produced, for a given fuel density, at a constant
temperature that gradually rises as a result of oxygen burning, until,
finally, a silicon-rich composition is produced. As a result of
turbulence, this oxygen layer, which we shall refer to as an oxygen
``flame'', is broadened and islands of nearly isothermal conditions
are produced. Here we flesh out those one-dimensional results in a
series of three-dimensional simulations.

In two previous three-dimensional studies,
\citet{Aspden08a,Aspden10} (henceforth Papers I and II),
turbulence-flame interactions in carbon-burning flames were examined
at small and large scales, respectively.  In Paper I, it was shown
that once the turbulence was sufficiently strong, the mixing of fuel
and heat was driven by turbulent mixing instead of thermal diffusion.
This resulted in a categorically different kind of flame, which was
referred to as a distributed flame. Paper II extended these
small-scale studies to (more realistic) larger length scales, where
scaling relations based on the theory of \cite{Damkohler40} were
predicted to reach a limiting behavior, resulting in a so-called
``$\lambda$-flame''.

In this paper, we focus on turbulence-flame interactions in the
trailing oxygen flame, which are expected to be significantly
different than in the carbon flame, due to the inductive nature of the
oxygen burning.  The specific question we address is whether
turbulence can lead to a greatly extended oxygen-rich
region that might have properties suitable for detonation.

\section{THEORETICAL DESCRIPTION}
\label{Sec:Theory}

We first recap the scaling relations for turbulent carbon flames from
Paper II, and then present theory describing possible modes of burning
in turbulent oxygen flames.  We fix the fuel to be carbon-oxygen at a
particular density and temperature, and then choose turbulent
conditions based on the flame properties of the fuel.
Since the oxygen burning time scale is determined by the temperature
resulting from carbon burning at the given density, the width of the
oxygen flame is most sensitive to that density and the turbulence
properties. For typical turbulent conditions in the supernova and an
initial composition of 40\% carbon and 60\% oxygen, \citet{Woosley10}
find that the density of greatest interest is $\rho_{12} = 2.5
\times 10^7$ g cm$^{-3}$ (we use $\rho$ and $T$ for total density
and temperature the suffix 12 denotes conditions before the carbon burns).  This gives a
post-carbon-flame temperature and density of
approximately $T_{16}=3.14\times 10^9$\,K and $\rho_{16}=1.69\times 10^7$\,g/cm$^3$,
respectively (here suffix 16 denotes conditions after carbon burning but before oxygen has burned).  Under these
conditions, the oxygen flame has an inductive burning
time scale of approximately $\tau_{16}=0.016$\,s, see \cite{Woosley10}.

Having fixed the fuel conditions, the two parameters
that can be varied are the rms turbulent velocity fluctuation
$\check{u}$ and the integral length scale $l$.  Turbulent premixed
flames are characterized through Karlovitz and Damk\"ohler numbers
\begin{equation}
\Ka^2_L=\frac{\check{u}^3l_L}{s_L^3l},
\quad\quad\mathrm{and}\quad\quad
\Da_L=\frac{s_Ll}{\check{u}l_L},
\end{equation}
where $s_L$ and $l_L$ are the laminar flame speed and width,
respectively.  These quantities represent the ratio of turbulent time scales 
at the Kolmogorov and integral length scales, respectively, and are
two dimensionless quantities that represent the parameter space.
As in Paper II, we focus on a fixed $\Ka_L$, corresponding to fixing
the energy dissipation rate $\varepsilon^*=\check{u}^3/l$ of the
turbulence in the star.  For fixed $\Ka_L$ and $\varepsilon^*$, 
$\check{u}=(\varepsilon^*l)^{1/3}$, it can be shown that 
$\Da_L\propto l^{2/3}$.  Therefore, the parameter space is
one-dimensional and can be represented equivalently by either 
$\Da_L$ or $l$. 

We assume that the Karlovitz number is constant and sufficiently high to
obtain a distributed carbon flame, see Paper I.  The turbulent carbon
flame properties will then depend on its turbulent nuclear burning time
scale $\tau_{12}^T$ (note the superscript $T$ differentiates the
turbulent from laminar burning time scales, again see Paper I), and
the properties of the turbulence, specifically the integral length
scale $l$ (recall that the turbulent intensity at fixed Karlovitz
number is determined by $\check{u}=(\varepsilon^*l)^{1/3}$).  Following
Paper II and \citet{Damkohler40}, by analogy with laminar flames, the
turbulent flame speed $s_{12}^T$ and width $l_{12}^T$ can be expressed
in terms of $\tau_{12}^T$ and a turbulent diffusion coefficient
$\mathcal{D}_T$ (not to be confused with the Damk\"ohler number $\Da$) as
\begin{equation}
s_{12}^T=\sqrt{\frac{\mathcal{D}_T}{\tau_{12}^T}},
\quad\quad\mathrm{and}\quad\quad
l_{12}^T=\sqrt{\mathcal{D}_T\tau_{12}^T},
\end{equation}
respectively.  These relations only hold when the time scale of the
turbulent eddies is shorter than the turbulent nuclear time scale of
the carbon fuel, i.e.\ for $\Da_{12}^T=\tau/\tau_{12}^T\ltaprx1$,
where $\tau$ is the turbulence time scale $\tau=l/\check{u}$.  Taking
a simple approximation $\mathcal{D}_T=\check{u}l$, the turbulent flame
speed and width were both shown to be proportional to $\Da_{12}^T$ when
$\Da_{12}^T\ltaprx1$.  For $\Da_{12}^T\gtaprx1$, the turbulence can
no longer broaden the flame, and the limiting $\lambda$-flame behavior
is reached (see Paper II), with local turbulent speed $s_{12}^\lambda$
and width $\lambda_{12}$ that depend on $\varepsilon^*$ and
$\tau_{12}^T$ only, according to the relations
\begin{equation}
s_{12}^{\lambda}=\sqrt{\varepsilon^*\tau_{12}^T}
\quad\quad\mathrm{and}\quad\quad
\lambda_{12}=\sqrt{\varepsilon^*{\tau_{12}^T}^3},
\end{equation}
respectively.  Note that $s_{12}^{\lambda}$ and $\lambda_{12}$ are
both constant.  The turbulent flame speed and width can therefore be
written as
\begin{equation}
\frac{s_{12}^T}{s_{12}^{\lambda}}=\frac{l_{16}^T}{\lambda_{12}}=
\left\{
\begin{array}{ll}
\Da_{12}^T & \mathrm{for\ }\Da_{12}^T\ltaprx1,\\
1&\mathrm{otherwise}.
\end{array}
\right.
\end{equation}

We now apply this theoretical approach to oxygen flames, where there
are three potential modes of burning: inductive, turbulent or
$\lambda$-flame.  Each mode will have a corresponding local flame
width and speed, which will be denoted $(l_{16},s_{16})$, 
$(l_{16}^T,s_{16}^T)$, and $(l_{16}^\lambda,s_{16}^\lambda)$, respectively.

The inductive mode is the simplest and is considered first.
In a frame of reference where the carbon flame is stationary, 
the incoming fluid speed is equal to the flame speed $u_0=s_{12}$.
The resulting oxygen flame has a width equal to $l_{16}=u_0\tau_{16}$, 
where the time taken for the oxygen to burn at a given density and 
temperature is $\tau_{16}$.  In the presence of turbulence, the
carbon flame speed is enhanced, but the oxygen flame remains 
slaved to carbon flame, and $l_{16}=u_0\tau_{16}$, only with 
$u_0=s_{12}^T$.  In the large-scale turbulence limit (see
\cite{Damkohler40,Peters99,Peters00}), the turbulent carbon flame
speed will be close to the turbulent intensity, and so we can take
$u_0=\beta\check{u}$, where $\beta$ should be expected to be order
unity, but can be as high as three, accounting for fluctuations and
the density jump across the carbon flame.  This defines
$s_{16}=\beta\check{u}$ and $l_{16}=\beta\check{u}\tau_{16}$ for 
a turbulent oxygen flame burning inductively.  

Defining Karlovitz and Damk\"ohler numbers for oxygen flames 
\begin{equation}
\Ka^2_{16}=\frac{\check{u}^3l_{16}}{s_{16}^3l},
\quad\quad\mathrm{and}\quad\quad
\Da_{16}=\frac{s_{16}l}{\check{u}l_{16}},
\end{equation}
reveals an interesting difference from carbon flames. 
Using $s_{16}=\beta\check{u}$, it can be shown that
$\beta^2\Ka_{16}^2\Da_{16}\equiv1$, which means that the parameter
space for oxygen flames is one-dimensional.
Therefore, under the assumption that the carbon flame is in the
large-scale turbulence limit (so $s_{16}=\beta\check{u}$), the
behavior of oxygen flames can be classified through the Damk\"ohler
number $\Da_{16}$ alone.  Note that for a fixed fuel, all of the relevant Damk\"ohler
numbers are constant multiples of each other,
e.g.\ $\Da_{16}=\sigma\Da_{12}^T$, where
$\sigma=\tau_{12}^T/\tau_{16}$ is a constant.  Also note that both
$s_{16}$ and $l_{16}$ can be shown to be proportional to
$\Da_{16}^{1/2}$.

If turbulent mixing can drive the flame, similar to the behavior in
Paper II, scaling relations for turbulent flame speed and width can be
predicted in terms of the oxygen burning time scale and the turbulent
diffusion coefficient $\mathcal{D}_T$ as
$s_{16}^T=\sqrt{\mathcal{D}_T/\tau_{16}}$ and 
$l_{16}^T=\sqrt{\mathcal{D}_T\tau_{16}}$.  These scaling relations
should only be expected to be possible for low values of the oxygen Damk\"ohler
number, i.e.\ $\Da_{16}=\ltaprx1$.  On the other hand,
for $\Da_{16}\gtaprx1$, it may be possible to produce
an oxygen $\lambda$-flame, where the flame speed and width would be
$s_{16}^{\lambda}=\sqrt{\varepsilon^*\tau_{16}}$ and
$\lambda_{16}=\sqrt{\varepsilon^*{\tau_{16}}^3}$, respectively.
As above, the turbulent oxygen flame speed and width can be predicted
to be
\begin{equation}
\frac{s_{16}^T}{s_{16}^{\lambda}}=\frac{l_{16}^T}{\lambda_{16}}=
\left\{
\begin{array}{ll}
\Da_{16} & \mathrm{for\ }\Da_{16}\ltaprx1,\\
1&\mathrm{otherwise}.
\end{array}
\right.
\end{equation}
The simulations in this study correspond to full-star conditions where
$\check{u}^*=2\times 10^7$\,cm/s on an integral length scale of
$L^*=1\times 10^6$\,cm, giving an energy dissipation rate of
$\varepsilon^*=8\times 10^{15}$\,cm$^2$/s$^3$, \citep[see][]{Roepke2007}.
Carbon fuel (40\%) at $\rho_{12}=2.5\times 10^7$\,g/cm$^3$ and $T_{12}=6\times 10^8$\,K
burns to $\rho_{16}=1.69\times 10^7$\,g/cm$^3$ and $T_{16}=3.14\times 10^9$\,K, which
has an inductive time scale for oxygen of $0.016$\,s.
This gives an oxygen $\lambda$-flame speed and width of
$s_{16}^\lambda=1.13\times 10^7$\,cm/s, and 
$\lambda_{16}=1.81\times 10^5$\,cm, respectively.

\begin{figure}
\centering
\plotone{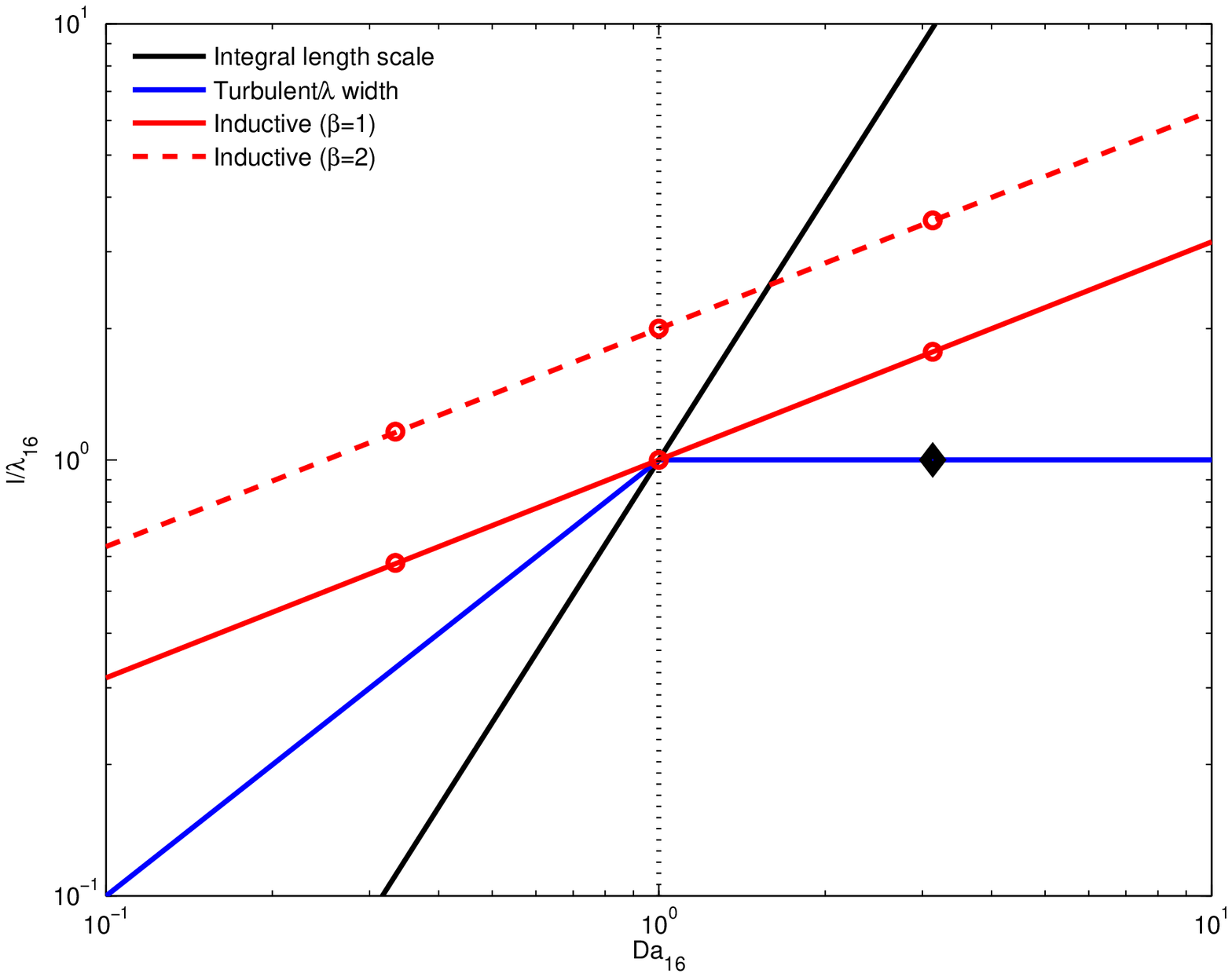}
\caption{Scaling relations for oxygen flame widths as a function of $\Da_{16}$.
The red lines show the inductive flame width $l_{16}$ (solid
for $\beta=1$, and dashed for $\beta=2$).  The blue line
shows the turbulent width $l_{16}^T$ if turbulent mixing drives the
flame, which scales with $\Da_{16}$ for $\Da_{16}\ltaprx1$ and is
equal to the $\lambda$-flame thickness for $\Da_{16}\gtaprx1$ .}
\label{Fig:LengthScaling}
\end{figure}

Figure~\ref{Fig:LengthScaling} depicts the scaling relations for
the different turbulent flame widths as a function of Damk\"ohler
number.  Recall $\Da_{16}\propto l^{2/3}$, so this can be thought of
as a function of integral length scale, shown by the thick black
line.  The red lines show the normalized inductive flame width
$l_{16}/\lambda_{16}$ (solid for $\beta=1$, and dashed for $\beta=2$).
The blue line shows the normalized turbulent width
$l_{16}^T/\lambda_{16}=\Da_{16}$ for ($\Da_{16}\ltaprx1$) and 
$l_{16}^T/\lambda_{16}=1$ for ($\Da_{16}\gtaprx1$), 
if turbulent mixing drives the flame.  The red circles correspond to
the simulations that will be considered in this study, specifically 
$\Da_{16}=1/3$, $1$, and $3$, with $\beta=1$ and $2$, and will be 
described in detail below.

The black diamond highlights the $\lambda$-flame thickness at 
$\Da_{16}=3$, which is smaller than the corresponding inductive
flame widths.  This means that the possibility of an oxygen
$\lambda$-flame is particularly interesting as it would be an example
of turbulence giving rise to a flame that is {\em narrower} than its
zero turbulence counterpart.  This is counterintuitive, as one expects
turbulence to {\em broaden} interfaces.  However, in this case,
turbulence acts to enhance heat transfer, allowing the flame to burn
more rapidly; turbulence can mix hot ash with cold fuel more rapidly
than the fuel is heated through inductive burning.

\section{SIMULATION DESCRIPTION}
\label{Sec:Numerics}

As in Papers~I and II, we use a low Mach number hydrodynamics code,
adapted to the study of thermonuclear flames, as described in
\citet{SNeCodePaper}.  The advantage of this method is that sound
waves are filtered out analytically, so the time step is set by 
the bulk fluid velocity and not the sound speed.  This is an enormous
efficiency gain for low speed flames.
We note that in all simulations presented here, the Mach number remains
below 0.1 (usually by an order of magnitude), and so compressibility 
effects are considered to be negligible.
The reactions rate here are taken from \citet{caughlan-fowler:1988} 
with screening.  The conductivities are those reported in 
\citet{timmes_he_flames:2000}, and the equation of state
is the Helmholtz free-energy based general stellar EOS described in
\citet{timmes_swesty:2000}.  We note that we do not utilize the
Coulomb corrections to the electron gas in the general EOS, as these
are expected to be minor at the conditions considered.

The non-oscillatory finite-volume scheme employed here permits the use
of implicit large eddy simulation (\iles).  This technique captures
the inviscid cascade of kinetic energy through the inertial range,
while the numerical error acts in a way that emulates the dissipative
physical effects on the dynamics at the grid scale, without the
expense of resolving the entire dissipation subrange.  An overview of
the technique can be found in \cite{GrinsteinBook07}.
\cite{Aspden08b} presented a detailed study of the technique using the
present numerical scheme, including a characterization that allowed
for an effective viscosity to be derived.  Thermal diffusion plays a
significant role in the flame dynamics, so it is explicitly included
in the model.  Species diffusion is significantly smaller, so
it is not included explicitly, but will be subject to numerical diffusion,
which can be considered to have an effective unity Schmidt number 
and exhibit the same behavior observed for viscosity in \cite{Aspden08b}.

The turbulent velocity field was maintained using the forcing term
used in Papers~I, II and \cite{Aspden08b}.  Specifically, a forcing
term was included in the momentum equations consisting of a
superposition of long wavelength Fourier modes with random amplitudes
and phases. The forcing term is scaled by density so that the forcing
is somewhat reduced in the ash. This approach provides a way to embed
the flame in a turbulent background, mimicking the much larger
inertial range that these flames would experience in a type Ia
supernova, without the need to resolve the large-scale convective
motions that drive the turbulent energy cascade.  \cite{Aspden08b}
demonstrated that the effective Kolmogorov length scale is
approximately $0.28\Delta x$, and the integral length scale is
approximately a tenth of the domain width.

\begin{figure}
\centering
\epsscale{0.8}
\plotone{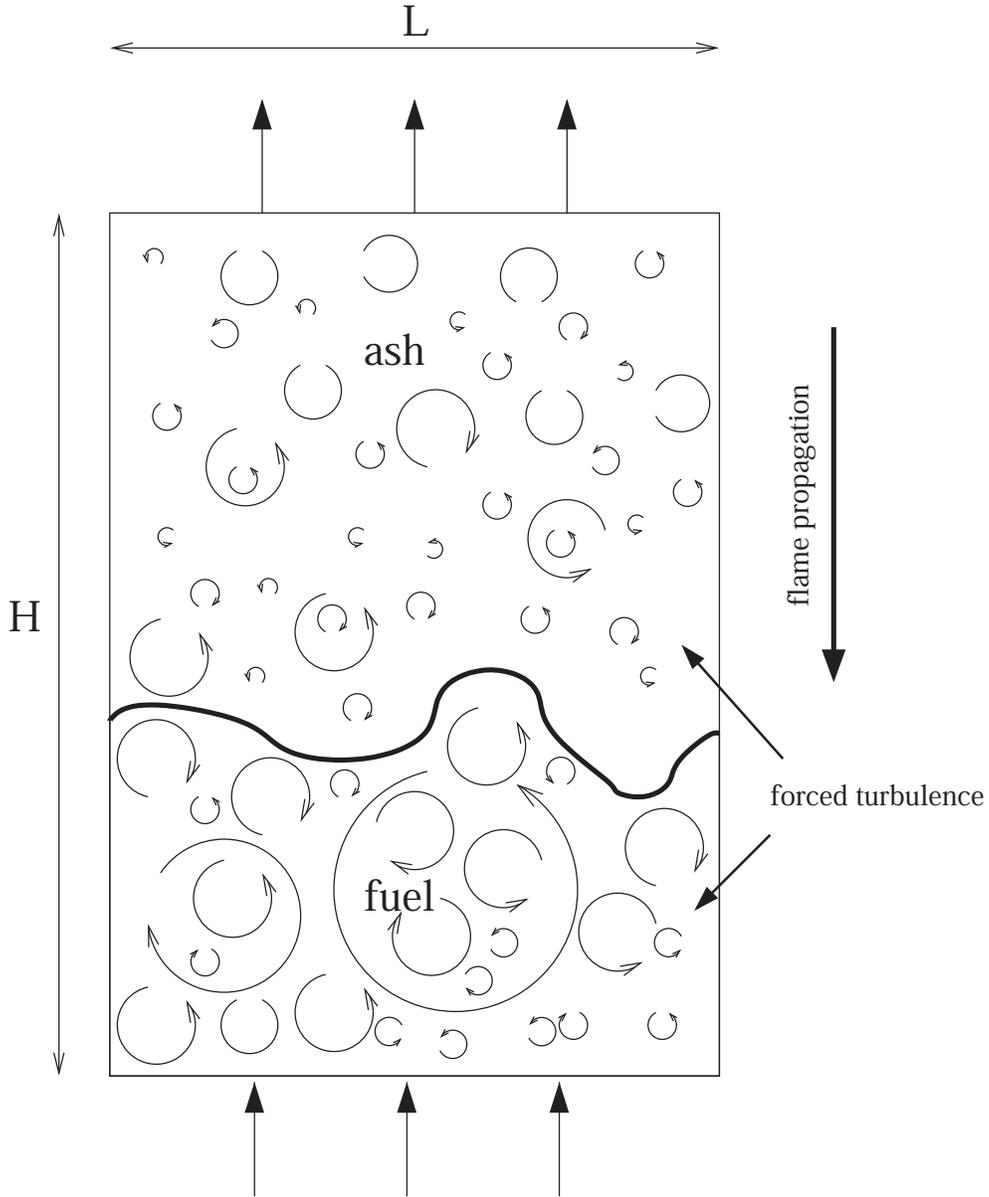}
\epsscale{1.0}
\caption{Diagram of the simulation setup (shown in two-dimensions for
clarity). The domain is initialized with a turbulent flow and a
flame is introduced into the domain, oriented to that the flame
propagates downwards against the imposed mean flow.  The turbulence is
maintained by adding a forcing term to the momentum equations. The top
and bottom boundaries are outflow and inflow, respectively. The side
boundaries are periodic.}
\label{Fig:Setup}
\end{figure}

Figure~\ref{Fig:Setup} shows the simulation setup.  The simulations
were initialized with oxygen fuel in the lower part of the domain and
sulphur ash in the upper part, resulting in a downward propagating flame.
A high-aspect ratio domain was used, with periodic lateral boundary
conditions, and outflow at the upper boundary.  Due to the huge
disparity in widths of carbon and oxygen flames, we are not able to
capture both accurately in the same simulation.  Therefore, we use carbon
post-flame conditions and only burn oxygen.  To recreate these
condition appropriately, specifically the warm oxygen in a post
carbon-flame mean velocity, we need to work in a moving frame of
reference.  Consequently, unlike papers I and II, a mean inflow was
specified at the lower boundary to replicate the desired conditions so
that an inductive oxygen flame could develop properly.  Using such an
inflow velocity in conjunction with the forcing term used to maintain
the turbulent velocity field requires some care.  It is possible to
use a turbulent inflow velocity, but we have opted not to take such an
approach.  Instead we specify a uniform inflow and use just the
forcing term itself to produce turbulence.  We found that this gave
satisfactory results, provided $\beta\ltaprx2$.

The conditions of particular interest are when $\Da_{16}\gtaprx1$.
Note from figure~\ref{Fig:LengthScaling} that the integral length
scale is expected to be much larger than all of the turbulent flame
widths under these conditions.  Given that the integral length scale
is approximately one-tenth of the domain width, this means that the
oxygen will burn extremely close to the inlet, and certainly before
the turbulence has become well-developed.  To account for this, the
inflow velocity, temperature and density were synthetically altered
(slower, cooler and more dense, respectively) to delay oxygen burning and ensure
that the flame burned approximately half-way through the domain, while
maintaining conditions close to the carbon post-flame.

Oxygen Damk\"ohler numbers of $1/3$, $1$ and $3$ were simulated to capture
the potential transition.  The aim is to detect the mode in which the
oxygen flame is burning, specifically, what is the local turbulent oxygen
flame width.  However, it is difficult to measure a local
turbulent flame width directly, and the widths from different modes of
burning may be difficult to distinguish.  Therefore, two inflow
velocities were used at each Damk\"ohler number, specifically,
$\beta=1$ and $\beta=2$.  This means that if the turbulent flame burns
inductively, the turbulent flame widths will differ by a factor of
approximately two.  However, if the flames are driven by turbulent
mixing, then the turbulent flame widths should be independent of
$\beta$.  This means that the crucial comparison required to determine
the burning mode is between flames at different inflow speeds at the 
same Damk\"ohler number, and direct measurements of local turbulent flame
widths do not need to be evaluated, nor are comparisons of these widths
necessary at different Damk\"ohler numbers.  

Simulations of one-dimensional zero-turbulence inductive flames and
three-dimensional inert turbulence were first obtained, and then
superimposed to initialize each calculation.  Each simulation was run
with a resolution of $256\times256\times1024$.  Adaptive mesh
refinement was not used.  Table \ref{Tab:SimProperties} gives the
conditions for the six simulations.

\section{RESULTS}
\label{Sec:Results}

\begin{figure}
\centering
\plottwo{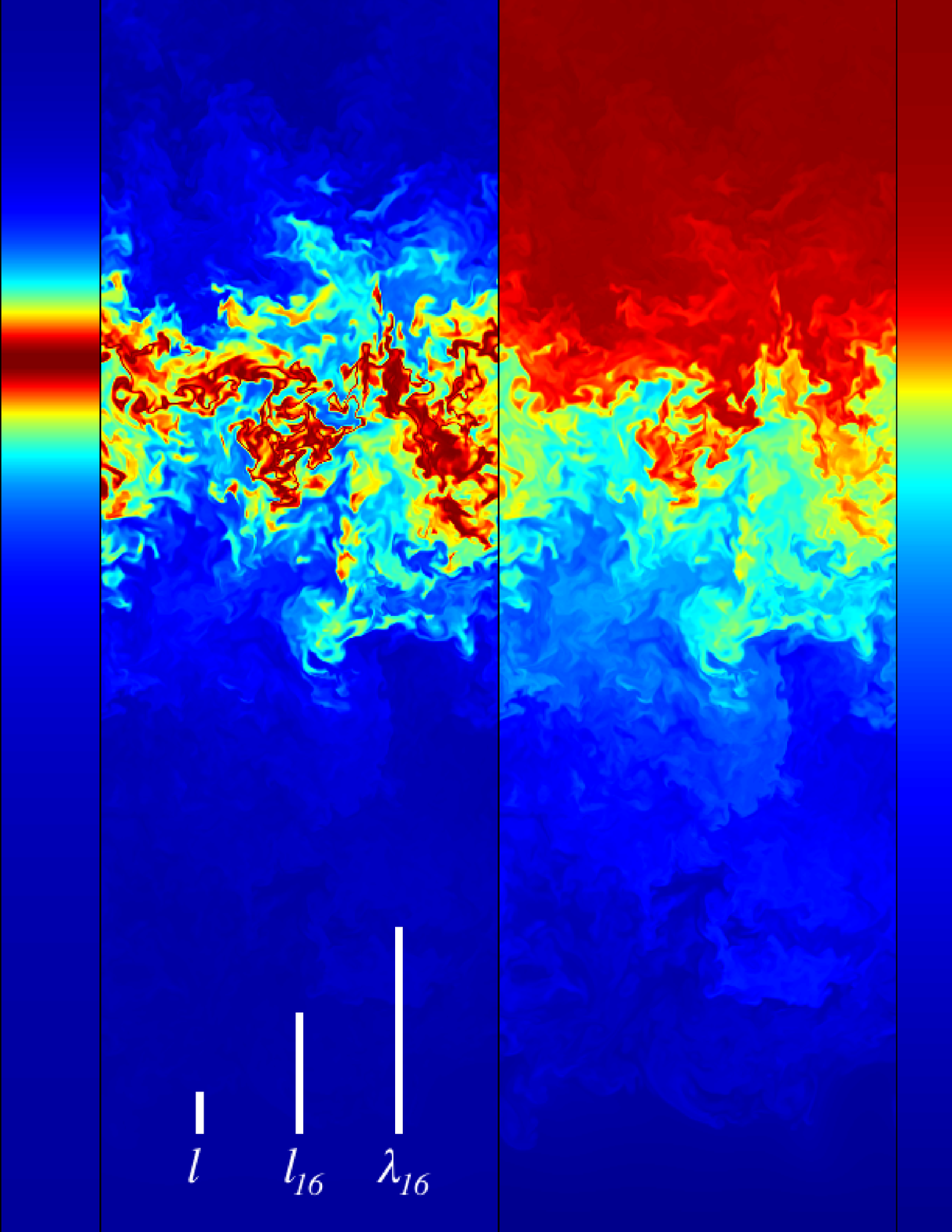}{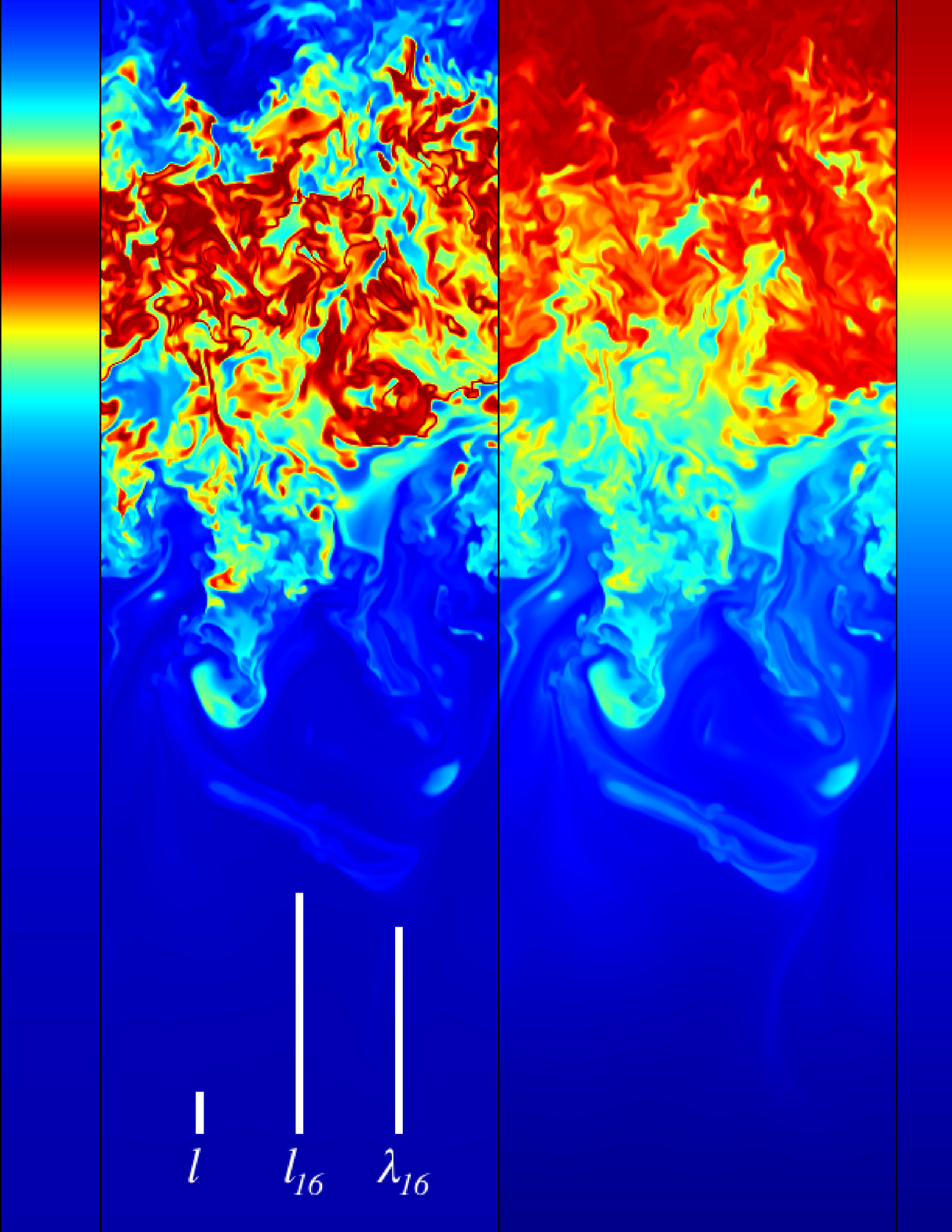}
\caption{Two-dimensional vertical slices (through three-dimensional
simulations) of burning rate (left) and temperature (right) for the
$\Da_{16}=1/3$ cases, (a) $\beta=1$ and (b) $\beta=2$. 
The two central panels show a snapshot of the turbulent 
simulations, and the narrow edge panels show the corresponding images
for the (zero turbulence) inductive flames for comparison.  
The white lines show three relevant length scales ($l$, $l_{16}$, and
$\lambda_{16}$).}
\label{Fig:SlicesDa03}
\end{figure}

Figures~\ref{Fig:SlicesDa03}(a,b) show two-dimensional vertical slices
(through three-dimensional simulations) of burning rate (left) and
temperature (right) for the $\Da_{16}=1/3$ cases, (a) $\beta=1$ and
(b) $\beta=2$.  The two central panels in each figure show a snapshot
of the turbulent simulations, and the narrow edge panels show the
corresponding images for the (zero turbulence) inductive flames for
comparison.  The white lines show three relevant length scales ($l$,
$l_{16}$, and $\lambda_{16}$).  Note that only $l_{16}$ differs between the
two cases (due to the dependence on $u_0$ and therefore $\beta$).
Warm oxygen is being fed from below at $u_0=\beta\check{u}$, burns to
sulphur, and leaves the domain through the top boundary.  In both
cases, there is a large volume of fuel burning, many times the
integral length scale.  It is clear that there is a high level of
turbulent mixing, but the width of each flame appears to be roughly
the same as the laminar inductive flame at the corresponding inflow speed.  Importantly, the $\beta=2$ flame
appears significantly broader than for $\beta=1$.  This suggests that
for $\Da_{16}\ltaprx1$, the flames burn inductively.

\begin{figure}
\centering
\plottwo{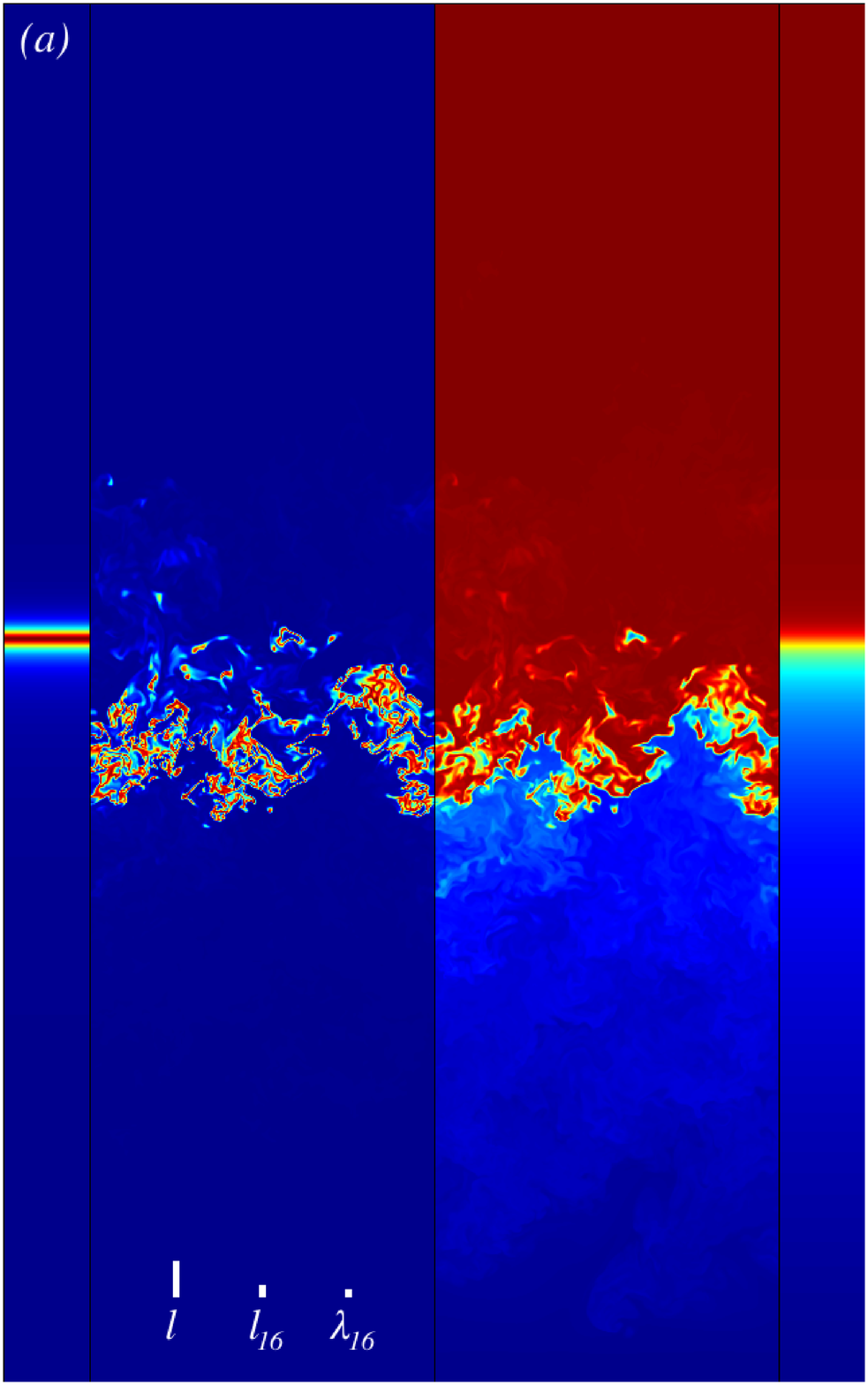}{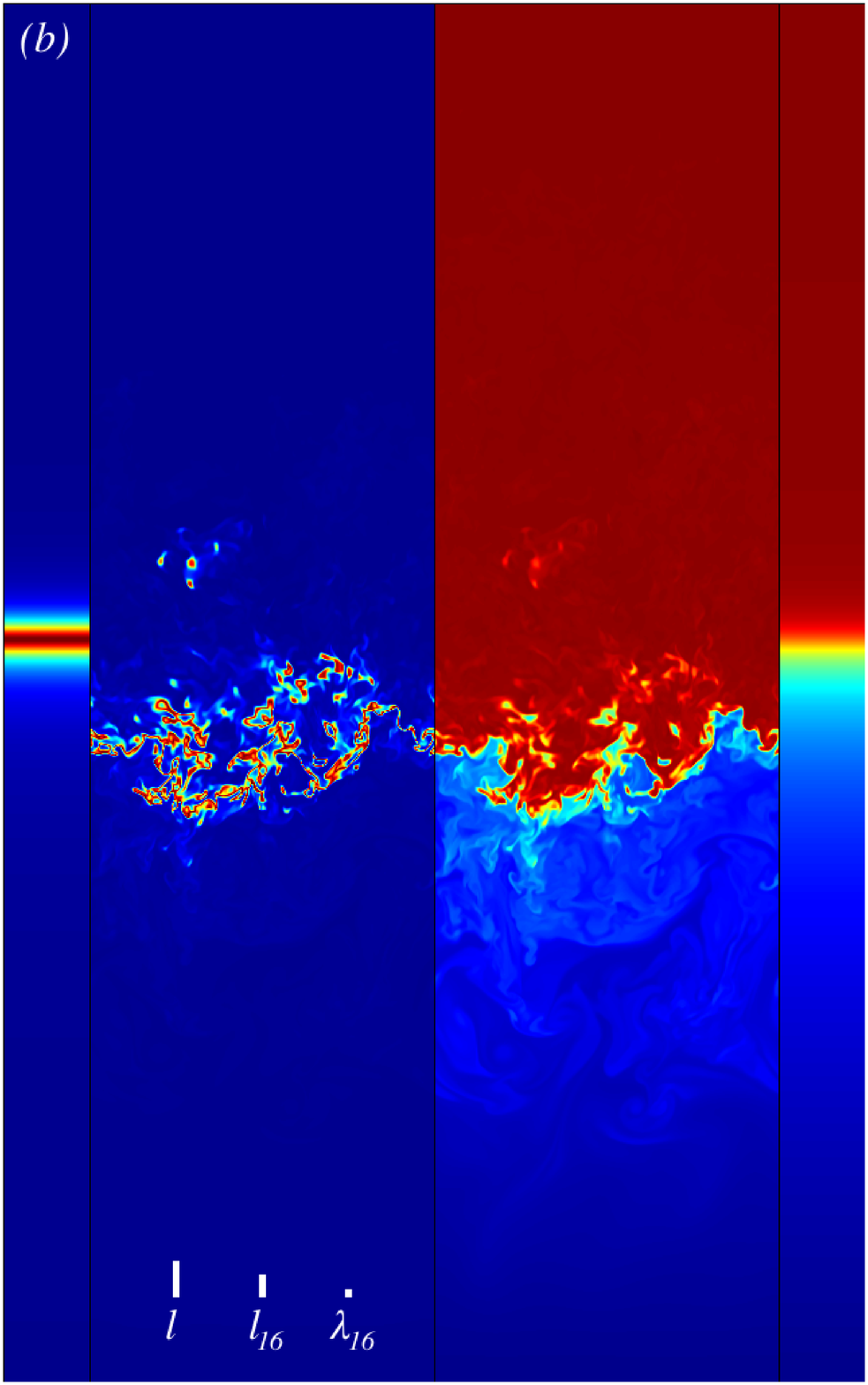}
\caption{As figure \ref{Fig:SlicesDa03} for the $\Da_{16}=3$ cases.}
\label{Fig:SlicesDa30}
\end{figure}

Figures~\ref{Fig:SlicesDa30}(a,b) show the corresponding slices for
the $\Da_{16}=3$ cases (the $\Da_{16}=1$ cases present intermediate
behavior, and are not shown).  Note the domain size and integral
length scale are 27 times larger than figure \ref{Fig:SlicesDa03}, and
the inflow and turbulent intensity is 3 times higher.
Correspondingly, the inductive length $l_{16}$ and $\lambda$-width are
relatively shorter here.  In both cases, the local turbulent flame
width appears to be narrower than the corresponding inductive flames
shown, and is even filamentary in places.  The crucial point to note
here is that the local turbulent flame width does not differ
significantly between the $\beta=1$ and $\beta=2$ cases.  Therefore,
turbulent mixing must be driving the flame propagation.  If the flames
were burning inductively, the $\beta=2$ case should be broader than
the $\beta=1$ case.  This is evidence that oxygen burns as a
$\lambda$-flame for $\Da_{16}>1$.

\begin{figure}
\centering
\plotone{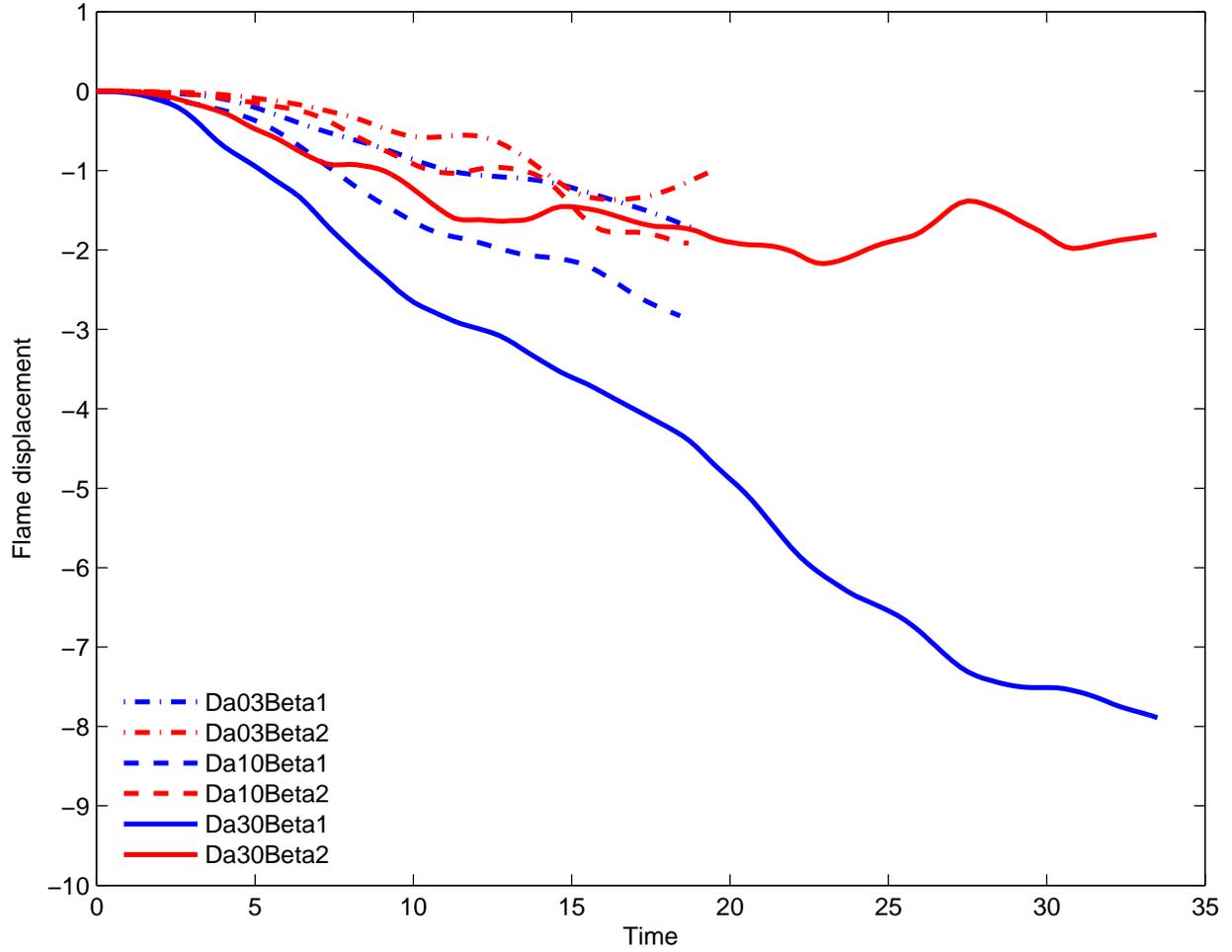}
\caption{Flame displacement from the initial position as a function 
of time for all six cases.  The displacement has been normalized by
the integral length scale $l$, and  the time has been normalized by
the integral length eddy turnover time $\tau=l/\check{u}$.}
\label{Fig:Displacement}
\end{figure}

An underlying assumption of the scaling analysis in
section~\ref{Sec:Theory} is that the nuclear time scale $\tau_{16}$ is
constant.  This assumption, combined with a turbulent flame width narrower than
$l_{16}$, suggests that the turbulent flame speed is faster than
$s_{16}$ (i.e.\ $u_0=\beta\check{u}$).  Figure~\ref{Fig:Displacement}
shows the flame displacement $\xi(t)=z_0(t)-z_0(0)$ from the initial
position $z_0(0)$ as a function of time for all six cases, where the
flame position has been defined as
\begin{equation}
z_0(t)=\frac{1}{A\left(\rho X_{16}\right)_0}\int_V \rho X_{16} \, \D V.
\end{equation}
The displacement has been normalized by the integral length scale $l$, and 
the time has been normalized by the integral length eddy turnover time 
$\tau=l/\check{u}$.  It is clearly evident that the Da30B1 flame is indeed 
burning significantly faster than $u_0$ and propagates towards the inflow 
boundary; we note that it is the high Damk\"ohler case (Da30B1) that is of 
particular interest.

This is an interesting consequence of turbulence-driven oxygen flame
propagation, because it suggests that the conventional idea that the
oxygen flame is locked at some distance (depending solely on $u_0$ and
$\tau_{16}$) behind the carbon flame \citep[e.g.][]{timmeswoosley1992,lisewski-distributed} cannot be the case under these
conditions.  For an order-of-magnitude analysis, assume the turbulent burning 
time scale of carbon at $\rho=2.5\times 10^7$ is of the order of 10$^{-3}$\,s,
which means the carbon $\lambda$-flame local speed and width are
approximately $s_{12}^\lambda\approx3\times 10^6$\,cm/s and
$\lambda_{12}\approx3\times 10^3$\,cm, respectively.  The
corresponding values for the resulting oxygen flame are
$s_{16}^\lambda=1.13\times 10^7$\,cm/s and
$\lambda_{16}=1.81\times 10^5$\,cm, respectively.  The oxygen
$\lambda$-flame is about an order of magnitude faster and two orders
of magnitude thicker than the carbon flame.  This suggests that the
flame actually burns as a single compound carbon-oxygen flame with
local flame speed and width close to that of the oxygen
$\lambda$-flame.

\begin{figure}
\centering
\plotone{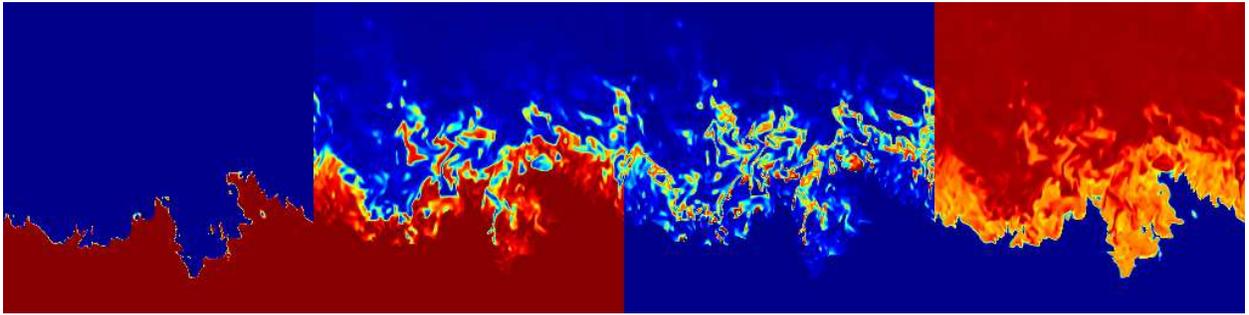}
\caption{Two-dimensional slices through a three-dimensional simulation
  of a compound carbon-oxygen flame.  The panels are carbon mass
  fraction, oxygen mass fraction, oxygen burning rate and temperature,
  respectively.}
\label{Fig:COFlame}
\end{figure}

Figure~\ref{Fig:COFlame} shows two-dimensional slices through a
three-dimensional simulation of a compound carbon-oxygen flame where
the inflow velocity and turbulent intensity were matched to case
Da30B1.  We emphasize that at this resolution, the carbon flame is far
from being well-resolved.  The simulation was initialized with a
discontinuity halfway up the domain, with cold fuel under hot ash.
Specifically, the fuel was at a density and temperature of
$\rho=2.5\times 10^7$\,g/cm$^3$ and $T=6\times 10^8$\,K, and consisted of 40\%
carbon and 60\% oxygen.  The initial ash consisted of 40\% magnesium
and 60\% sulphur, with density and temperature of $\rho\approx
1.39\times 10^7$\,g/cm$^3$ and $T\approx 3.8\times 10^9$\,K, respectively, and
was allowed to evolve to the appropriate state.  The figure panels are
carbon mass fraction, oxygen mass fraction, oxygen burning rate and
temperature, respectively.  It appears that, as expected, turbulent
mixing is able to drive a compound flame.

\begin{figure}
\centering
\plotone{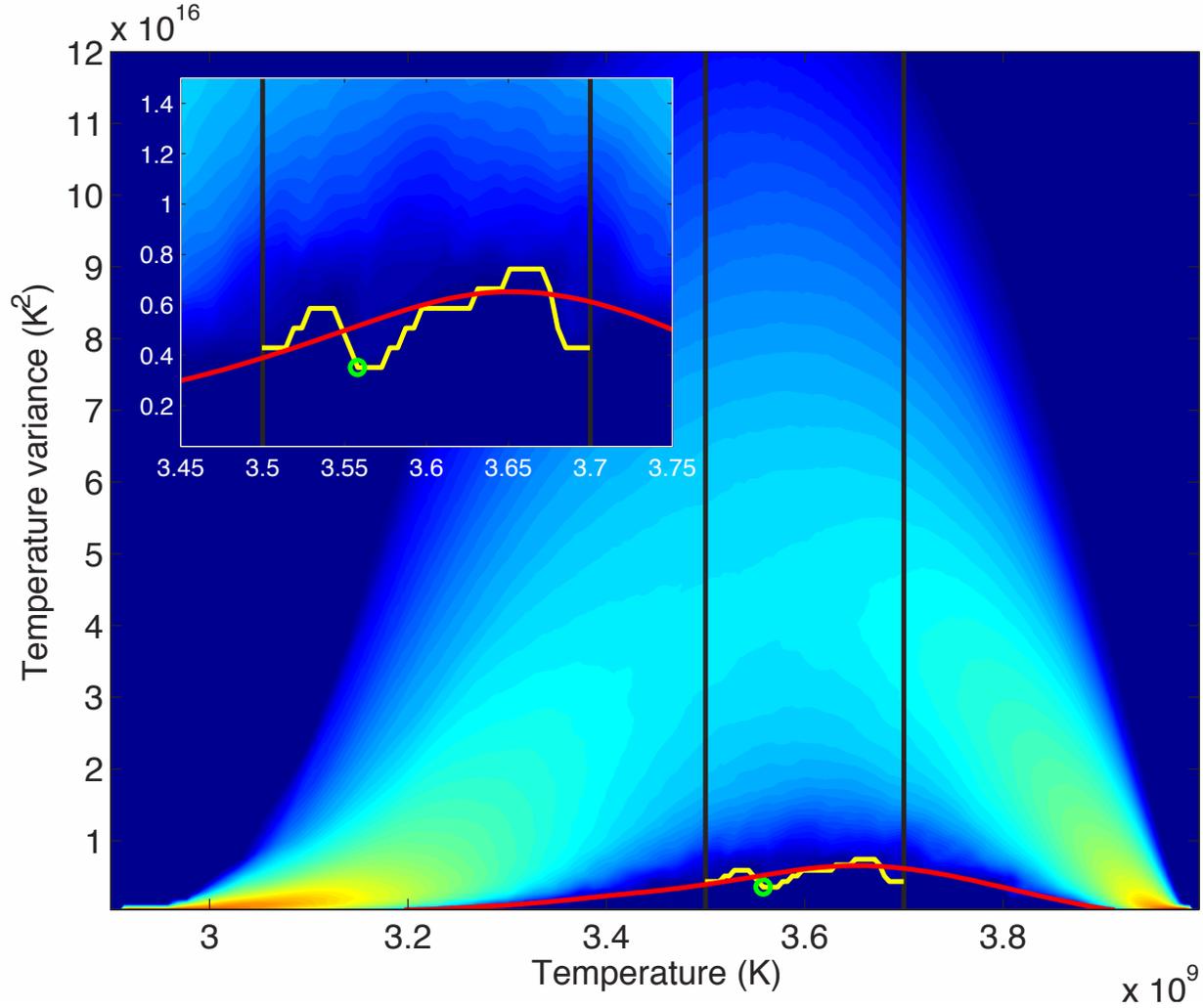}
\caption{Joint probability density function of temperature and 
  temperature variance evaluated using a 2.33\,km cubic filter 
  for case Da10B2.  The red curve denotes the zero turbulence
  case filtered in the same way.  The yellow line denotes the 
  minimum variance achieved for each temperature in the range
  denoted by vertical black lines, over which a potential
  transition to detonation was proposed by \cite{Woosley10}.
  The green circle denotes the lowest variation that was found
  within this temperature range.  The inset shows a zoom of
  the conditions of interest.}
\label{Fig:PdfTempVar}
\end{figure}

The potential transition to detonation suggested by \cite{Woosley10}
requires the formation of a region of approximately 10\,km at a temperature of 
approximately $3.6\times10^9$\,K.  To investigate the existence
of such a region, the temperature field from the Da10B2 case averaged
using a top-hat cubic filter of size $33^3$ cells. which corresponds to a length 
for each side
of approximately 2.33\,km.  The filtered temperature and temperature variance were 
found at each point in space and time, and are plotted in the form of a 
joint probability density function in figure \ref{Fig:PdfTempVar}.
The red curve denotes the zero turbulence
case filtered in the same way.  The yellow line denotes the 
minimum variance achieved for each temperature in the range
denoted by vertical black lines, over which a potential
transition to detonation was proposed by \cite{Woosley10}.
Almost every point within the flame lies above the laminar profile,
but there are low probability events that have a low variance
within the required temperature range.  However, the minimum
variance over this range (denoted by the green circle) is 
approximately $3.5\times10^{15}$\,K$^2$.  
The candidate case from figure 3b of \cite{Woosley10}
has a temperature of approximately $3.67\times10^9\pm2.5\times10^7$\,K.  
Assuming a uniform distribution gives a variance of approximately 
$2\times10^{14}$\,K$^2$, which is an order of magnitude less than in the 
current simulation.   Allowing for a larger range, say $\pm$1e8\,K, 
gives a variance of approximately $3\times10^{15}$\,K$^2$.
These estimates are not much lower than the laminar flame (shown by
the red curve), which suggests that the turbulence does not lead to 
the formation of a plateau or ledge under these conditions.
Furthermore, the filter size here is much smaller than required by
about a factor of 4, and the variance will only increase with a 
larger filter size.  This does not provide support for the suggests
transition to detonation in oxygen for the conditions studied, but
it should be noted that the simulations in \cite{Woosley10} had a 
turbulent power over 15 times that used here.

\section{CONCLUSIONS}
\label{sec:conclusions}

The theoretical treatment of distributed carbon-burning thermonuclear
flames from \cite{Aspden10} has been applied to the trailing oxygen
flames and compared with three-dimensional simulations over a range of
Damk\"ohler numbers.  It was shown that for $\Da_{16}\ltaprx1$,
turbulence does not greatly alter the flame from one in
which the oxygen burns purely inductively.  Since turbulence
accelerates the carbon flame however, the width of the oxygen flame
is enormously broader than in the laminar case. 
For $\Da_{16}\gtaprx1$, turbulence enhances heat transfer and drives
flame propagations that is {\em narrower} than the corresponding
zero turbulence inductive oxygen flame.  This is somewhat 
counterintuitive as turbulence typically broadens interfaces 
rather than sharpening them.  A consequence of burning in this 
limit is that the oxygen can burn faster than the inductive flame speed
(but is limited by the carbon flame speed of course). 
Therefore, the oxygen flame does not trail behind the carbon flame 
(at a distance equal to the post carbon-flame velocity times the oxygen 
burning time scale), but burns as a compound carbon-oxygen.
This suggests that a single level set is a suitable flame model for the
compound flame under these conditions.  Averaging the temperature field
using a cubic filter suggested that the temperature variance in at
the desired conditions is too high to support the potential transition
to detonation in oxygen proposed in \cite{Woosley10}.
However, this does not preclude this kind of transition under different
conditions, such as higher turbulence or lower densities, and it 
should be borne in mind that only one such event would be required,
and in the star there are many realizations.

\acknowledgements

A.~J.~A.\ and J.~B.~B.\ were supported by the Applied Mathematics Research
Program of the U.S. Department of Energy under Contract No. DE-AC02-05CH11231.
At UCSC this research has been supported by the NASA Theory
Program NNX09AK36G and the DOE SciDAC Program (DE-FC02-06ER41438).
The computations presented here were performed on the ATLAS Linux
Cluster at LLNL as part of a Grand Challenge Project. 

\bibliographystyle{apj}
\bibliography{rt}

\clearpage

\clearpage

\clearpage

\begin{table}
\begin{tabular}{|l||c|c|c|c|c|c|}
\hline
Case & Da03B1 & Da03B2 & Da10B1 & Da10B2 & Da30B1 & Da30B2 \\
\hline
Damk\"ohler number ($\Da_{16}$)  & 1/3 & 1/3 & 1 & 1 & 3 & 3 \\ 
Inflow factor ($\beta$)  & 1 & 2 & 1 & 2 & 1 & 2 \\
Domain width ($L$) [km] & $3.50$  & $3.50$  & $18.1$  & $18.1$  & $100$  & $100$ \\
Domain height ($H$) [km] & $14.0$  & $14.0$  & $72.4$  & $72.4$  & $400$  & $400$ \\
Integral length scale ($l$) [km] & $0.350$  & $0.350$  & $1.81$  & $1.81$  & $10.0$  & $10.0$ \\
Turbulent intensity ($\check{u}$) [km/s]  & $65.4$  & $65.4$  & $113$  & $113$  & $200$  & $200$ \\
Inflow velocity ($u_0^\dagger$) [km/s] & $61.46$  & $12.65$  & $10.05$  & $20.68$  & $15.73$  & $34.90$ \\
Inflow density ($\rho_0^\dagger$) [$\times10^7$\,g/cm$^3$] & $1.786$  & $1.739$  & $1.895$  & $1.846$  & $1.998$  & $1.940$ \\
Inflow temperature ($T_0^\dagger$) [$\times10^9$\,K] & $2.950$  & $3.035$  & $2.790$  & $2.870$  & $2.656$  & $2.740$ \\
\hline
\end{tabular}
\caption{Simulation properties.  The $\dagger$ denotes the synthetic
  inflow conditions that were chosen to position the laminar inductive
  flame close to the half-way point in the domain.  The desired
  conditions were $\rho_0=1.69\times 10^7$\,g/cm$^3$, $T_0=3.14\times 10^9$\,K and
  $u_0=\beta\check{u}$.}
\label{Tab:SimProperties}
\end{table}

\clearpage

\clearpage

\clearpage

\clearpage

\clearpage

\end{document}